\documentclass[amsmath,pre,twocolumn,nofootinbib,showpacs]{revtex4}
\usepackage{bm,amssymb,graphicx}
\DeclareMathAlphabet{\varmathbb}{U}{bbold}{m}{n}

\newcommand{\dlangle}{{\langle\!\langle}}
\newcommand{\drangle}{{\rangle\!\rangle}}
\newcommand{\bfone}{{\bm 1}}

\newcommand{\bfA}{\bm{A}}
\newcommand{\bfP}{\bm{P}}

\newcommand{\bfQ}{\bm{Q}}

\newcommand{\notD}{{{\not}D}}
\newcommand{\bfepsilon}{{\hat{\bm \epsilon}}}
\newcommand{\eqn}[1]{Eq.\ (\ref{Eqn:#1})}
\newcommand{\eqns}[1]{Eqs.\ (\ref{Eqn:#1})}
\newcommand{\eqnref}[1]{(\ref{Eqn:#1})}
\newcommand{\pp}[2]{\ensuremath{\frac{\partial #1}{\partial #2}}}
\newcommand{\opp}[2]{\ensuremath{\frac{d #1}{d #2}}}

\begin{document}
\title{Efficient computation of the first passage time distribution of the generalized master equation by steady-state relaxation}
\author{David Shalloway}
\email{dis2@cornell.edu}
\affiliation{Biophysics Program, Dept. of Molecular Biology and Genetics, Cornell University, Ithaca, New York  14853, USA}
\author{Anton K.\ Faradjian}
\affiliation{Department of Physics, Cornell University, Ithaca, New York  14853, USA}
\begin{abstract}
The generalized master equation or the equivalent continuous time random walk equations can be used to compute the macroscopic first passage time distribution (FPTD) of a complex stochastic system from short-term microscopic simulation data.  The computation of the mean first passage time and additional low-order FPTD moments can be simplified by directly relating the FPTD moment generating function to the moments of the local FPTD matrix.  This relationship can be physically interpreted in terms of steady-state relaxation, an extension of steady-state flow.
Moreover, it is amenable to a statistical error analysis that can be used to significantly increase computational efficiency.
The  efficiency improvement can be extended to the FPTD itself by modelling it using a Gamma distribution or rational function approximation to its Laplace transform.

\end{abstract}
\maketitle

\newpage

\section{Introduction}
The first passage time distribution (FPTD) \cite{VanKampen:92} concisely describes the kinetics of macroscopic transitions of complex macromolecular systems; e.g., the transition of a disordered heteropolymer from random to specifically-absorbed conformations \cite{Golumbfskie:99} or the dynamics of protein folding \cite{Lee:03a,Lee:03b}. If a statistical ensemble of systems is prepared at time $t=0$ in an initial metastable macroscopic state (macrostate) $i$ and $P_f(t)$ is the probability that an ensemble member is in absorbing final macrostate $f$ at time $t$, then the FPTD is
\begin{equation}
\label{Eqn:varphi}
\varphi(\tau) = \left. dP_f(t)/dt\right|_{t=\tau} \;.
\end{equation}
We assume that $f$ is the only absorbing state and that the system is ergodic, so
\begin{equation}
\label{Eqn:normalization}
\int_0^\infty \varphi(\tau) d \tau = P_f(\infty) = 1 \;.
\end{equation}
The mean first passage time (MFPT) is the first moment $\langle \tau \varphi \rangle \equiv \dlangle \tau \drangle$, where
\begin{eqnarray*}
\langle x \rangle & \equiv & \int_0^\infty x(\tau) \, d\tau  \,, \\
\dlangle x \drangle & \equiv & \int_0^\infty x(\tau) \varphi(\tau) \, d\tau  \,.
\end{eqnarray*}
It determines the transition rate as $\dlangle \tau \drangle^{-1}$ \cite{Hanggi:90,Reimann:99}.

Except for the simplest models, $\varphi(\tau)$ and its moments can not be analytically computed.  Moreover, direct numerical computation, e.g. by molecular or stochastic dynamics, is often unaffordable.  For example, proteins typically have $10^3-10^5$ conformational degrees-of-freedom and the macroscopic timescales of interest can be $>10^{12}$ times larger than the microscopic timescale \cite{Brooks:88}, so huge amounts of computational effort would be needed for direct simulation.

Coarse-graining can overcome these problems. The essential idea is to subdivide the macroscopic transition into a network of discrete intermediate mesoscopic transitions that are fast enough for feasible computation (e.g., using Monte Carlo \cite{Berry:95} or molecular dynamics \cite{Faradjian:04} methods) yet slow enough (relative to the microscopic timescale) for approximation by a first-order stochastic equation.

The simplest approximation of this sort is a Markovian master equation \cite{VanKampen:92} for $\bfP(t)$, the $N$-vector that specifies ensemble probability over the intermediate mesoscopic states and initial and final states (see \cite{Despa:05} and references therein for examples in the context of protein folding).  However, this approach will only be accurate when each mesoscopic transition can be characterized as a simple Poisson process. This is not the case for many important problems because of fractal or quasi-diffusive dynamics or because the timescale that would be needed to achieve the Markovian limit is too long for direct computation \cite{Sadana:01,Yang:03}.

The non-Markovian classical generalized master equation is a more robust approximation that can be used in this situation \cite{Kenkre:74,Zwanzig:83,gme_note}.  Assuming injection at $t=0$ of ensemble members into initial state $i$, it is
\begin{equation}
\label{Eqn:generalized}
\opp{\bfP(t)}{t}  = \delta(t-0^+) \bfepsilon_i -\int_0^\infty \Gamma(\tau) \cdot \bfP(t-\tau) d\tau \;,
\end{equation}
where $\Gamma(\tau)$ is the $N \times N$ matrix of transition functions which include memory effects and $\bfepsilon_s$  denotes the basis vector that has component $s$ equal to 1 and all other components equal to 0.
The first term, along with the boundary condition $\bfP(-\infty)=0$, initializes the system with no memory at $t=0$. Conservation of probability and causality imply that
\[
 \bfone \cdot \Gamma(\tau) =  0 \;, \quad
\Gamma_{s' s}(\tau) \le 0 \quad (s' \ne s)\;,
\]
(where $\bfone$ is the $N$-vector with all components equal to 1) and $s$ and $s'$ take values corresponding to $i$, $f$, and all the intermediate mesoscopic states. Therefore
\begin{equation}
\label{Eqn:conservation_of_probability}
\bfone \cdot d \bfP(t)/dt = \delta(t-0^+) \;.
\end{equation}
Since $f$ is absorbing,
\begin{equation}
\label{Eqn:irreversible}
\Gamma (\tau) \cdot \bfepsilon_f = 0  \;.
\end{equation}
If $\Gamma(\tau)$ can be analytically computed (e.g., by projection \cite{Zwanzig:61,Kubo:78}), then \eqns{varphi} and \eqnref{generalized} can be used to compute $\varphi(\tau)$.

When $\Gamma(\tau)$ can not be computed, an approach using numerical
simulations can be employed  \cite{Faradjian:04}.  By initializing
multiple simulations in state $s$ and determining the distribution
of waiting times for first-transitions to the other mesoscopic
states, simulations can be used to determine $K(\tau)$, the $N
\times N$ \emph{local FPTD matrix} (sometimes called the
``first-jump waiting time'' matrix). $-K_{s'  s}(\tau) \; (s \ne
s')$ is the probability density that, after arriving at state $s$, a
system waits for an interval $\tau$ before first leaving and that it
goes to $s'$.  For compact notation, we define the diagonal elements
of $K(\tau)$ as $K_{s s}(\tau) \equiv -\sum_{s' \ne s} K_{s'
s}(\tau)$. Thus, like $\Gamma(\tau)$, $K(\tau)$ satisfies
\begin{equation}
\label{Eqn:K_properties}
\bfone \cdot K(\tau) =  0 \;, \quad K_{s s'}(\tau) \le 0 \quad (s' \ne s), \quad K(\tau) \cdot \bfepsilon_f  =  0 \;,
\end{equation}
and, by its definition and the assumption that $f$ is the only absorbing state,
\begin{equation}
\label{Eqn:K_integral}
\int_0^\infty K_{s s}(\tau) \, d \tau =  1 \quad (s\ne f) \;.
\end{equation}

$K(\tau)$ is the kernel of an alternative representation of stochastic dynamics with memory---the generalized continuous time random walk (CTRW).  Originally introduced to describe random walks on lattices \cite{Montroll:65,Haus:87}, the CTRW was generalized \cite{Scher:73, Klafter:80} to a form  that can be extended to memory-dependent stochastic processes on a mesoscopic network of arbitrary connectivity.  In our notation and assuming $t=0$ initialization in state $i$ this is
\begin{subequations}
\label{Eqn:QK}
\begin{eqnarray}
\label{Eqn:QKa}
\opp{\bfP(t)}{t} & = & \bfQ(t) - \int_0^\infty K_D(\tau) \cdot \bfQ(t-\tau) d \tau \\
\label{Eqn:QKb}
\bfQ(t) & = &  \delta(t-0^+) \bfepsilon_i + \int_0^\infty K_\notD (\tau) \cdot \bfQ(t-\tau) d \tau \;,
\end{eqnarray}
\end{subequations}
where $K_D$ and $K_\notD$ are the matrices comprised, respectively, of the diagonal or off-diagonal elements of $K$, and the boundary conditions are $\bfP(-\infty) = \bfQ(-\infty)=0$.   \eqn{QKb} implies that $Q_s(t) dt$ is the probability that an ensemble member makes a transition to state $s$ within the interval $[t, t + dt)$, and \eqn{QKa} states that $d\bfP(t)/dt$ is the difference between the incoming and outgoing probability flows.

\eqns{generalized} and \eqnref{QK} provide equivalent descriptions of the temporal evolution of $\bfP(t)$ \cite{Zwanzig:83,Kenkre:74}, and comparing their Laplace transforms shows that
\begin{subequations}
\begin{eqnarray}
\label{Eqn:Laplace}
\widetilde{\Gamma}(u) & = & u \widetilde{K}(u)\cdot [I- \widetilde{K}_D(u)]^{-1} \\
\widetilde{K}(u) & = & \widetilde{\Gamma}(u) \cdot [ uI + \widetilde{\Gamma}_D(u)]^{-1} \,,
\end{eqnarray}
\end{subequations}
where $I$ is the identity matrix and we denote the Laplace transform of $g(u)$ as $\tilde{g}(u) \equiv \int_0^\infty e^{-u \tau} g(\tau) \, d \tau$. However, even though \eqn{Laplace}  determines $\widetilde{\Gamma}(u)$ from $\widetilde{K}(u)$, the inverse Laplace transform needed to determine $\Gamma(\tau)$ can be difficult, if not impossible, to compute. Thus, even though \eqns{generalized} and \eqnref{QK} are formally equivalent, only the CTRW formulation provides a practical way to use mesoscopic numerical simulations to compute $\varphi(\tau)$.

Faradjian and Elber \cite{Faradjian:04} have recently demonstrated the feasibility of integrating \eqns{QK} with  $K(\tau)$ determined by molecular dynamics  to compute transitions along a single reaction-coordinate.  However, computing $\varphi(\tau)$ by this approach is computationally wasteful since it is determined to high temporal resolution even though the experimentally relevant information is usually contained in only a few of its low-order moments. The unnecessary price paid is that the complete functional form of $K(\tau)$ must be determined by many (expensive) numerical simulations.

In Sec.\ \ref{Sec:moment_generating_function} we present a
fundamental new relationship between the FPTD moments and those of
$K(\tau)$ or $\Gamma(\tau)$. We show in Sec.\ \ref{Sec:heuristic}
that this relationship can be understood as an extension of the
steady-state flux-over-population method \cite{Farkas:27,Hanggi:90}
of computing rate constants to the case of \emph{steady-state
relaxation}. In Sec.\ \ref{Sec:efficient} we show that combining
this relationship with statistical error analysis yields a more
accurate and efficient computational algorithm.  In Sec.\
\ref{Sec:modelling} we demonstrate how $\varphi(\tau)$ can be
modelled using a few of its moments and a Gamma distribution or a
rational function approximation to its Laplace transform.

\section{The FPTD Moment Generating Function}
\label{Sec:moment_generating_function}
The Laplace transform of $\varphi(\tau)$ gives the FPTD moment generating function:
\begin{equation}
\label{Eqn:generating_function}
\sum_{k=0}^\infty \frac{\dlangle \tau^k \drangle \alpha^k}{k!}= \tilde{\varphi}(-\alpha) \;.
\end{equation}
We assume, as is true in most cases of interest, that $\Gamma_{s' s}(\tau)$ $(\forall \; s,\, s')$ and $\varphi(\tau)$ decays faster than $e^{- \alpha_{\rm max} \tau}$ as $\tau \to \infty$ for some positive $\alpha_{\rm max}$ (corresponding to the slowest process in the system):
\begin{equation}
\label{Eqn:convergence}
\lim_{\tau \to \infty} \Gamma_{s' s}(\tau) e^{\alpha_{\rm max} \tau}= \lim_{\tau \to \infty} \varphi(\tau) e^{\alpha_{\rm max} \tau}  =  0 \;.
\end{equation}
Thus, $\tilde{\varphi}(-\alpha)$ is analytic in a neighborhood  about 0 and can be differentiated to yield all moments. (This assumption is not essential, but simplifies the discussion.  If it is not true, the analysis will still be valid for the finite moments.)

Taking the Laplace transform of \eqn{varphi} gives
\begin{equation}
\label{Eqn:varphi_P}
\tilde{\varphi}(-\alpha) = -\alpha \bfepsilon_f \cdot \widetilde{\bfP}(-\alpha)  \;.
\end{equation}
\eqn{normalization} implies that $\bfepsilon_f \cdot \widetilde{\bfP}(-\alpha)$ is not analytic at $\alpha = 0$, so expanding $\tilde{\varphi}(-\alpha)$ using this form is delicate \cite{singular_note}.   To avoid this inconvenience, we use \eqn{conservation_of_probability} to rewrite \eqn{varphi} in a form that does not explicitly involve $P_f(t)$:
\begin{eqnarray}
\varphi(\tau) & = & \delta(\tau-0^+)- \bfone \cdot \Pi \cdot \left. d\bfP(t)/dt\right|_{t=\tau}  \nonumber \\
\label{Eqn:varphi_projected}
& = & \delta(\tau-0^+)- \bfone \cdot \left. d\bar{\bfP}(t)/dt\right|_{t=\tau} \,,
\end{eqnarray}
where
\[
\Pi \equiv I - \bfepsilon_f \otimes \bfepsilon_f
\]
is the projection operator into the \emph{dynamic subspace} of non-absorbing states $s \ne f$ and we use the notation
\begin{eqnarray*}
\bar{\bfA} & \equiv & \Pi \cdot \bfA \\
\bar{M} & \equiv & \Pi \cdot M \cdot \Pi
\end{eqnarray*}
to denote projected vectors $\bar{\bfA}$ and matrices $\bar{M}$.
\eqn{varphi_projected} relates the FPTD to the loss of probability from the dynamic states.  Its Laplace transform is
\begin{equation}
\label{Eqn:varphi_Pbar}
\tilde{\varphi}(-\alpha) =  1 + \alpha \bfone \cdot \widetilde{\bar{\bfP}}(-\alpha) \;.
\end{equation}
This form is advantageous because \eqns{varphi}, \eqnref{normalization}, \eqnref{conservation_of_probability}, and \eqnref{convergence} imply that
\begin{equation}
\label{Eqn:bar_convergence}
\lim_{t \to \infty} \bar{\bfP}(t) \, e^{\alpha_{\rm max} t}= 0 \,,
\end{equation}
so $\widetilde{\bar{\bfP}}(-\alpha)$ is analytic at $\alpha=0$.

To complete the solution, we express $\bar{\bfP}$ in terms of $\widetilde{\bar{\Gamma}}$ by projecting \eqn{generalized} [using $\Pi \cdot \Gamma \cdot \bfP =  \bar{\Gamma} \cdot \bar{\bfP}$, which follows from \eqn{irreversible}] and taking its Laplace transform to get
\[
u \widetilde{\bar{\bfP}}(u) = \bfepsilon_i - \widetilde{\bar{\Gamma}}(u) \cdot \widetilde{\bar{\bfP}}(u) \;.
\]
The solution  is
\begin{equation}
\label{Eqn:P_bar_u}
\widetilde{\bar{\bfP}}(u) = [u\Pi + \widetilde{\bar{\Gamma}}(u)]^{-1} \cdot \bfepsilon_i\,,
\end{equation}
where the use of the matrix pseudo-inverse (i.e., the inverse within the dynamic subspace) is implied here and below.  Combining this with \eqn{varphi_Pbar} gives
\begin{equation}
\label{Eqn:Gamma_Laplace_relationship}
\tilde{\varphi}(-\alpha) =  1-\alpha \bfone \cdot [ \alpha \Pi-\widetilde{\bar{\Gamma}}(-\alpha)]^{-1} \cdot \bfepsilon_i
\end{equation}
The right-hand-side is analytic at $\alpha=0$ because $\widetilde{\bar{\Gamma}}(0)$ is invertible within the dynamic subspace.

Using \eqn{Laplace}, we reexpress this in terms of $\widetilde{\bar{K}}$:
\begin{equation}
\label{Eqn:K_Laplace_relationship}
\tilde{\varphi}(-\alpha)  =  -\bfone \cdot \widetilde{\bar{K}}(-\alpha) \cdot [\Pi + \widetilde{\bar{K}}_\notD (-\alpha)]^{-1} \cdot \bfepsilon_i \;.
\end{equation}
Since $\widetilde{K}(-\alpha)$ is analytic at $\alpha = 0$ even without projection, \eqn{K_Laplace_relationship} can equivalently be written in unprojected form as \cite{Elber_note}
\[
\tilde{\varphi}(-\alpha)  =  \bfepsilon_f \cdot [I + \widetilde{K}_{{\not} D} (-\alpha)]^{-1} \cdot \bfepsilon_i  \;.
\]

Equations \eqnref{Gamma_Laplace_relationship} and
\eqnref{K_Laplace_relationship} provide the fundamental relationship
between the FPT moment generating function and $\Gamma$ and $K$.

\section{The moment generating function and steady-state relaxation}
\label{Sec:heuristic}
To elucidate the physical significance of \eqn{Gamma_Laplace_relationship}, we compare the computation of the MFPT using the generalized master equation with the computation of the transition rate (${\rm MFPT}^{-1}$) using the steady-state flux-over-population method \cite{Farkas:27,Hanggi:90}. The latter computes the rate  as the magnitude of the flux of systems divided by the total dynamic population in a steady-state situation.

We begin to relate the generalized master equation to the steady-state by noting that the solution of \eqn{generalized}, $\bfP(t)$, gives the response of a linear system to an impulse and so is the Green's function for the general solution: If systems are injected continuously at a non-negative rate $r(t)$ beginning at $t=0$, the resultant population distribution vector $\bfP[r;t]$ will satisfy
\begin{equation}
\label{Eqn:P_r_GME}
\opp{\bfP[r;t]}{t} = \theta(t) r(t) \bfepsilon_i - \int_0^\infty \Gamma(\tau) \cdot \bfP[r;t-\tau] d\tau
\end{equation}
with boundary condition $\bfP[r,-\infty] = 0$ ($\theta$ is the Heaviside step-function).  This has the solution
\begin{equation}
\label{Eqn:P_r}
\bfP[r;t] = \int_0^t \bfP(t-t') r(t') d t' \;.
\end{equation}

Unlike the Green's function $\bfP(t)$, which satisfies $\bfone \cdot \bfP(t) =1 \; (t>0)$, the general solution $\bfP[r;t]$ is unnormalized; the total  population is $\bfone \cdot \bfP[r;t] = \theta(t) \int_0^t r(t') dt'$, which can increase without bound as $t \to \infty$ because of the accumulation of systems in $f$. To avoid this complication, we follow the approach used above and focus on the projected dynamic population vector $\bar{\bfP}[r;t]$, which is bounded.

The steady-state case corresponds to the asymptotic ($t \gg \alpha^{-1}_{\rm max}$) regime with $r(t) = j$, where $j$ is a positive constant.  More generally, we consider \emph{steady-state relaxation}: the asymptotic regime with $r(t) = j \exp(-\alpha t)$ ($\alpha \ge 0$).  \eqns{bar_convergence} and \eqnref{P_r} imply the asymptotic form
\[
\bar{\bfP}[j e^{-\alpha t};t] \sim j e^{-\alpha t} \bar{\bfP}_\infty[\alpha] \qquad (\alpha \ge 0 \,,t \gg \alpha_{\rm max}^{-1})  \;,
\]
where  $\bar{\bfP}_\infty[\alpha]$ is a vector constant. Substituting this into the projected form of \eqn{P_r_GME}, multiplying by $\exp(\alpha t)/j$ and taking the limit $t \to \infty$ gives
\[
- \alpha \bar{\bfP}_\infty[\alpha] = \bfepsilon_i - \int_0^\infty e^{\alpha \tau} \bar{\Gamma}(\tau) d \tau \cdot \bar{\bfP}_\infty[\alpha]
\]
with solution
\[
\bar{\bfP}_\infty[\alpha] = -[\alpha \Pi-\widetilde{\bar{\Gamma}}(-\alpha)]^{-1} \cdot \bfepsilon_i \,,
\]
where the pseudo-inverse is again implied.
Comparing this with \eqn{P_bar_u} implies that
\[
\widetilde{\bar{\bfP}}(-\alpha)  =  \bar{\bfP}_\infty[\alpha]\,,
\]
so \eqn{varphi_Pbar} implies that
\begin{equation}
\label{Eqn:varphi_Pinfty}
\varphi(-\alpha)  = 1+ \alpha \bfone \cdot \bar{\bfP}_\infty[\alpha] \;.
\end{equation}

We see that the Laplace transform of $\bar{\bfP}(t)$, and hence the FPTD generating function, is simply related to the steady-state relaxation dynamic population vector.  The steady-state calculation of the transition rate is a special case of this more general relationship:  \eqns{generating_function} and \eqnref{varphi_Pinfty} imply that
\begin{equation}
\label{Eqn:MFPT_steadystate}
\dlangle \tau \drangle = \bfone \cdot \bar{\bfP}_\infty[0] \;.
\end{equation}
Since $j \bar{\bfP}_\infty[0]$ is the steady-state solution for  constant flux $j$ and its inner product with $\bfone$ is the sum over the population in all the dynamic states, \eqn{MFPT_steadystate} states that the MFPT is the total dynamic population over the flux. This is equivalent to the statement \cite{Farkas:27,Hanggi:90} that the transition rate is the flux-over-(dynamic) population.
\section{Efficient calculation of the FPTD moments}
\label{Sec:efficient}

\eqns{generating_function} and \eqnref{Gamma_Laplace_relationship} imply that $\langle \varphi \rangle = 1$, in agreement with \eqn{normalization}. Expanding \eqn{Gamma_Laplace_relationship} to first order gives
\begin{equation}
\label{Eqn:first_moment}
\dlangle \tau \drangle = \bfone \cdot \widetilde{\bar{\Gamma}}(0)^{-1} \cdot \bfepsilon_i = \bfone \cdot \langle \bar{\Gamma} \rangle ^{-1} \cdot \bfepsilon_i\;.
\end{equation}
The same result would be obtained if we ignored all memory effects and approximated $\Gamma(\tau) \approx \delta(\tau) \langle \Gamma \rangle$, which is equivalent to replacing the generalized master equation with a regular master equation having $\Gamma = \langle \Gamma \rangle$. Differences between the moments of these two equations only appear in higher order.

When only $K(\tau)$ is known, we can use \eqn{Laplace} or expand \eqn{K_Laplace_relationship} to reexpress \eqn{first_moment} in terms of $\bar{K}$:
\begin{equation}
\label{Eqn:K_first_moment}
\dlangle \tau \drangle = \bfone \cdot \langle \tau \bar{K}_D\rangle \cdot \langle \bar{K} \rangle ^{-1} \cdot \bfepsilon_i  \;.
\end{equation}
To use this relationship to compute $\dlangle \tau \drangle$ from numerical simulation data, the time-averages on its right-hand-side can be approximated by
\begin{subequations}
\label{Eqn:estimating}
\begin{eqnarray}
\label{Eqn:estimate_tau}
\langle \tau \bar{K}_D \rangle_{s s}& \approx & n_s^{-1} \sum_{i=1}^{n_s} \tau^s_i \,, \\
\label{Eqn:naive}
\langle \bar{K}\rangle_{s' s} & \approx & n_{s' s}/n_s
\end{eqnarray}
\end{subequations}
where $n_s$ is the number of simulations that were initiated in state $s$, $n_{s's}$ is the number of those simulations that made their first transition to state $s'$ and $\{ \tau^s_i: i=1, \ldots,n_s\}$ is the set of first transition times for the simulations initiated in $s$.  This result is much easier to compute than numerically solving the CTRW \eqns{QK} and then integrating \eqn{varphi} to compute $\dlangle \tau \drangle$.  Moreover it does not introduce quantization error, as occurs when numerically solving \eqns{QK}; its estimate for $\dlangle \tau \drangle$ equals that which would be obtained using the CTRW equations in the limit where the numerical quantization size $h \to 0$.
To prove this, note that we have already proved that \eqn{K_first_moment} and the CTRW calculation are equivalent when the exact $K(\tau)$ and moments are used.  These exact values would be obtained in the limit of an infinite amount of simulation data. The result obtained with a finite amount of simulation data can be viewed as an approximation to the exact result.  Alternatively, it can be viewed as the exact result for the problem in which $K_{s's}(\tau)$ is proportional to a sum of $\delta$-functions, each corresponding to one of the waiting times in the set $\{\tau^{s's}_i:i=1,\ldots,n_{s's}\}$ of numerically computed local first passage times from $s$ to $s'$ i.e., $K_{s's}(\tau) = -n_s^{-1} \sum_{i=1}^{n_{s's}} \delta[\tau-\tau^{s's}_i]$.  The numerical $K(\tau)$ computed in the limit $h \to 0$ and the moments of $K$ computed using \eqns{estimating} are both exact for this modified problem, and thus must yield the same result.

\subsection{Improving accuracy and efficiency by sampling adjustment}
The accuracy of the steady-state computation of $\dlangle \tau \drangle$ will depend on the quality of the statistical estimation of $\langle \tau \bar{K}_D \rangle$ and $\langle \bar{K}\rangle$ provided by \eqns{estimating}.  The simplest way to use these equations would be to follow the procedure used to estimate $K(\tau)$ in the CTRW approach and to initiate the same number of simulations in each state; i.e., $n_s=n_{\rm tot}/(N-1)$, where $n_{\rm tot}$ is the total number of simulations to be performed. (The denominator is $N-1$ because no simulations are initiated in the final state.) However, this procedure is not optimal because it does not account for differences in the sensitivity of the result to errors in different states.  For example, the inverse matrix $\langle \bar{K} \rangle^{-1}$ appearing in \eqn{K_first_moment} can be particularly sensitive to errors in small matrix elements corresponding to bottlenecks in the probabilistic flow where there tend to be fewer transitions in the ``forward'' direction.  Since the expected root-mean-square (rms) statistical errors of the matrix elements $\langle \bar{K} \rangle_{s's}$ are inversely proportional to $n_s$, overall accuracy will be improved if $n_s$ is increased for the bottleneck states while being decreased for other states to keep $n_{\rm tot}$ constant.

We can use \eqn{K_first_moment} to analyze the dependence of $\sigma^2$, the variance of $\dlangle \tau \drangle$, on the $n_s$ and thereby to quantitatively optimize effort allocation. To simplify notation we define
\begin{eqnarray*}
p_{s'}^s & \equiv &  -\langle \bar{K}_{s's}\rangle \,, \\
\phi_s & \equiv & \langle \tau \bar{K}_D \rangle_{ss}  \,.
\end{eqnarray*}
The $p_{s'}^s$ are multinomial probabilities governing first transitions out of state $s$, which by \eqns{K_properties} and \eqnref{K_integral} satisfy $\sum_{s' \ne s} p^s_{s'} = 1$.  Accounting for the reduction in the standard error of the mean resulting from repeated sampling, making the approximation that the statistical errors in $\phi_s$ are independent of those in the $p_{s'}^s$ \cite{independent}, and using the propagation of errors formula, we estimate
\begin{equation}
\label{Eqn:variance}
\sigma^2 =  \sum_{s \ne f} n_s^{-1}\left[ \left(\pp{\dlangle \tau \drangle}{\phi_s}\right)^2  \sigma^2_{\tau;s} + \sum_{s'',s' \ne s }
\pp{\dlangle \tau \drangle}{p_{s''}^s} \sigma^2_{s;s'' s'} \pp{\dlangle \tau \drangle}{p_{s'}^s} \right]
\end{equation}
where
$\sigma^2_{\tau;s}$ is the variance of the $\{\tau^s_i\}$ and
\[
\sigma^2_{s;s'' s'} = \left\{ \begin{array}{cc} p_{s'}^s [1-p_{s'}^s] & (s''=s') \\
                                                -p_{s''}^s p_{s'}^s   & (s'' \ne s')
                               \end{array} \right. \quad (s'',s' \ne s)
\]
is the approximate multinomial variance tensor for state $s$ \cite{binomial_simplification}. Since the cost of a simulation is proportional to its duration, the expected cost of $n_s$ simulations initiated at state $s$ will be $n_s  \phi_s$.  Minimizing $\sigma^2$ with respect to the $n_s$ while maintaining a constant total cost implies the optimality conditions
\begin{equation}
\label{Eqn:optimal}
n_s = c\, \phi_s^{-1/2} \sqrt{
\left(\pp{\dlangle \tau \drangle}{\phi_s}\right)^2 \sigma^2_{\tau;s} + \sum_{s'',s' \ne s }
\pp{\dlangle \tau \drangle}{p_{s''}^s} \sigma^2_{s;s'' s'}  \pp{\dlangle \tau \drangle}{p_{s'}^s} } \,
\end{equation}
where $c$ is a constant chosen so that $\sum_{s \ne f} n_s \phi_s= \texttt{cost}$.
\eqn{optimal} determines the $n_s$ as explicit functions of the $\phi_s$, $\sigma^2_{\tau;s}$, and $p_{s'}^s$.  To estimate these parameters, we can first perform a pilot run with a small number of simulations for each state.
More simulations can then be added to the pilot simulations so that the combined set satisfies \eqns{optimal} \cite{adaptive}.  \eqn{variance} can be used to estimate the error of the final result computed with the combined set of simulations to determine if the accuracy goal has been met.

Empirically, we have found that efficiency can be further improved to a small extent by replacing the maximum likelihood estimator of the $p_{s'}^s$ used in \eqn{naive} by a Bayes-Laplace estimator (Appendix A). This estimator was used in the example discussed below but only gave noticeable improvement for the low-accuracy (e.g., 25--50\%) results \cite{used_Bayes}.

\subsection{Example}
\label{Sec:example}
We compared the efficiency of the sampling-adjusted steady-state procedure with that of the standard CTRW procedure using the two-dimensional entropic barrier model studied by Faradjian and Elber \cite{Faradjian:04}. They computed the FPTD for transitions under Brownian dynamics with potential energy function $U(x,y)= x^6 + y^6 \exp(-100 x^2) [1-\exp(-100 y^2)]$ from an initial state with $x=-1$ to a final state with $x=0.714$ at $kT$= 0.5 and friction coefficient $\gamma=0.1$  The ``exact'' value of $\dlangle \tau \drangle$ was computed using the CTRW method with five linearly-ordered intermediate states and $n_s =5,000$ numerical simulations initiated at each state (i.e., a total of $n_{\rm tot}= 30,000$ simulations were used with $\texttt{cost}= \texttt{maxcost}= 5,000 \sum_{s \ne f}\phi_s$).  To assess the accuracy of the method as cost was decreased, we used their simulation data to determine the geometric rms error \cite{geometric_rms} of the CTRW estimates of $\dlangle \tau \drangle$ when fewer simulations were used corresponding to $\texttt{maxcost}/\texttt{cost} =$ 2, 4, 8, 16, 32, 64, 128, and 256.  For each value of $\texttt{cost}$
CTRW estimates of $\dlangle \tau \drangle$ were computed for 4,000 random data subsets, and their geometric rms error was computed relative to the ``exact'' value \cite{CTRW_computation}.

To assess the performance of the sample-adjusted steady-state procedure for a specified $\texttt{cost}$, we first performed a pilot run (with $n_s$ the same for all states) costing 1/4  $\texttt{cost}$, used the estimated values of $\phi_s$, $\sigma^2_{\tau;s}$, and $p_{s'}^s$ and \eqns{optimal} to optimize the distribution across states of additional simulations costing 3/4 $\texttt{cost}$, and evaluated $\dlangle \tau \drangle$ using \eqns{K_first_moment}, \eqnref{estimate_tau}, and the Bayes-Laplace proportions estimator (Appendix A) with the combined set of simulations. This procedure was repeated 4,000 times to estimate the geometric rms error.  Additional tests showed that the results were not highly sensitive to the size of the pilot run.

The geometric rms errors for both methods as a function of $\texttt{cost}$ are plotted in Fig.\ 1 and show that the sampling adjustment increased efficiency slightly more than two--fold. For example, $\texttt{cost} = \texttt{maxcost}/8$ was needed to achieve $\sim 12\%$ accuracy using the standard CTRW method, while only $\texttt{maxcost}/16$  was needed for $\sim 11\%$ accuracy with the adjusted steady-state method. Examination of the optimized $n_s$ showed that this gain occurred because a $\sim 4$--fold increase in the sampling frequency at a bottleneck caused a $\sim \sqrt{4}$--fold improvement in the associated dominating error.

The extent of sampling adjustment in this problem was limited because there were only five mesoscopic states among which effort could be reallocated.  Larger adjustments, and larger gains in efficiency, may be possible in larger problems if the increase in the number of mesoscopic states exceeds the relative increase in the number of bottlenecks. Such gains could be particularly important for very costly problems (e.g., those arising when studying protein conformational transitions).

Expressions analogous to \eqn{K_first_moment} for the higher FPT moments can be obtained by analytically expanding \eqn{K_Laplace_relationship} in terms of the $\langle \tau^k \bar{K} \rangle$ \cite{note2}. Although the optimal sampling conditions for simultaneously computing multiple moments differ from \eqns{optimal}, we expect that benefit will still be achieved even if sampling is adjusted using these equations. Of course, even better results will be obtained if the optimization analysis is extended to the multiple moment case.

\section{Efficient modelling of the FPTD}
\label{Sec:modelling}
In some cases we will need to compute not just the moments, but also $\varphi(\tau)$ to a low temporal resolution commensurate with experimental results.  We can extend the efficiency improvement obtained in the moment computations to this case by modelling the FPTD using a few of its low-order moments and an appropriate functional expansion.

\subsection{Modelling using the Gamma distribution}
A Gamma distribution of the form
\[
f(\beta,\gamma; \tau) = \frac{\beta (\beta \tau)^{\gamma-1} e^{-\beta \tau}}{\Gamma(\gamma)} \,,
\]
(where $\beta>0$, $\gamma>0$; here $\Gamma$ denotes the Euler Gamma function, not the transition matrix) provides a simple model.  It decays exponentially as $t \to \infty$, thereby matching the expected asymptotic behavior of $\varphi(\tau)$, and it is simple to choose $\beta$ and $\gamma$ so as to match the first two moments of the FPTD  \cite{exponential_fit}:
\begin{subequations}
\label{Eqn:Gamma_dist_fit}
\begin{eqnarray}
\langle \tau^k f \rangle & = & \dlangle \tau^k \drangle \qquad k = 1,2 \\
& \Rightarrow & \nonumber \\
\gamma & = & \frac{\dlangle \tau \drangle^2}{ \dlangle \tau^2 \drangle -\dlangle \tau \drangle^2} \\
\beta & = &  \gamma/\dlangle \tau \drangle
\end{eqnarray}
\end{subequations}
[The denominator of the expression for $\gamma$ is equal to $\dlangle (\tau - \dlangle \tau \drangle)^2 \drangle$, and so is guaranteed to be positive.]  Additional moments could be included by modelling  $\varphi(\tau)$ as a sum of Gamma distributions, but problems with non-unique parameter fitting can arise.

\subsection{Modelling using a rational function approximation to $\tilde{\varphi}$}
In some cases better results can be obtained by approximating  $\tilde{\varphi}(u)$ as a rational function
\begin{equation}
\label{Eqn:rational_function}
\tilde{\varphi}(u)  \approx  R_{m,n}(u) = \frac{1 + p_1 u + \ldots p_m u^m}{1 + q_1 u + \ldots q_n u^n} \qquad (n>m) \;.
\end{equation}
Here we have fixed the zeroth-order terms in the numerator and denominator so that $R_{m,n}(0) =\tilde{\varphi}(0)=1$, as required by \eqn{normalization}.  We require $n-m >1$ so that $\lim_{u \to \infty}R_{m,n}(u)$ vanishes at least as fast as $s^{-2}$, implying that its inverse Laplace transform will vanish at the origin, corresponding to $\varphi(0) = 0$.  Since the only singularities of $R_{m,n}(u)$ are poles, the inverse Laplace transform is easy to compute.  Moreover, if the only poles are on the negative real axis (not guaranteed), the inverse transform will be the sum of decaying exponentials, thereby providing a natural model for $\varphi(\tau)$.  We use this property as a validity check and do not accept (potentially overfitted) approximations that have poles off the negative real axis.

The $p_i$ and $q_i$ are fixed by requiring that $\tilde{\varphi}(u)=R_{m,n}(u) $ at $m+n$ non-zero values of $u_k$ ($k= 1, \ldots,m+n$).  To choose the $\{u_k \}$ appropriately we note that the most important structure of $\varphi(\tau)$  occurs at scale $\tau \sim \dlangle \tau \drangle$.  Therefore, the important structure of $\tilde{\varphi}(u)$ will occur at scale $u \sim \dlangle \tau \drangle^{-1}$.  Thus we choose $u_k  = k/\dlangle \tau \drangle$ and require
\begin{eqnarray}
\label{Eqn:rational_matching}
\tilde{\varphi}(k/\dlangle \tau \drangle)=  \dlangle e^{-k\tau/\dlangle \tau \drangle} \drangle & = & R_{m,n}(k/\dlangle \tau \drangle) \\
&& (k = 1,\ldots, m+n) \,.\nonumber
\end{eqnarray}
The statistical error of the exponential moments grows as $k$ increases because the exponential will down-weight a larger fraction of the data points. This limits the accuracy of the ``high-frequency'' components of the moment-modelled $\varphi(\tau)$ to be the same as that of the directly integrated $\varphi(\tau)$ \cite{denser_interp}.

The first moment of $\varphi$ obtained using \eqn{rational_function} will be close to, but will not exactly match the MFPT. An exact match can be obtained by replacing \eqn{rational_function} with the constrained rational function
\begin{eqnarray}
\tilde{\varphi}(u) & \approx &  R_{m,n}^{\dlangle \tau \drangle}(u) \nonumber \\
& = & \frac{1 + p_1 u + \ldots p_m u^m}{1 + (\dlangle \tau \drangle + p_1) u + q_2 u^2 + \ldots q_n u^n}
\label{Eqn:constrained_rational_function} \\
&&\qquad\qquad \qquad \qquad \qquad \qquad(n>m)\,. \nonumber
\end{eqnarray}
This satisfies
\begin{equation}
\label{Eqn:derivative}
\left. \frac{d R_{m,n}^{\dlangle \tau \drangle}(u)}{du} \right|_{u=0} = -\dlangle \tau \drangle \;,
\end{equation}
so the first moment of its inverse Laplace transform will exactly equal the MFPT.  The constraint on the derivative of $R_{m,n}^{\dlangle \tau \drangle}$ replaces the use of the $k=1$ constraint in \eqn{rational_matching}, so when \eqn{constrained_rational_function} is used we only match $R_{m,n}^{\dlangle \tau \drangle}(u_k)= \tilde{\varphi}(u_k)$ for $k= 2, \ldots, m+n$ \cite{constrained_rational}. In most cases the estimates obtained using \eqns{rational_function} or \eqns{constrained_rational_function} will be similar.
The two lowest-order approximations of this type are those involving the MFPT and either one or two exponential moments corresponding to approximating $\varphi(\alpha)$ as $R^{\dlangle \tau \drangle}_{0,2}(\alpha)$ or $R^{\dlangle \tau \drangle}_{0,3}(\alpha)$.

We illustrate the method using the two-dimensional model discussed above.  In Fig.\ 2 we compare the FPTD computed by Faradjian and Elber  \cite{Faradjian:04} using the CTRW method with the approximated FPTD's computed using the MFPT and either zero, one or two exponential moments.  The single-exponential fit obtained using the MFPT alone [i.e., $\exp(-t/\dlangle \tau \drangle)$] misses much important detail, but a fairly good representation is obtained by additionally  matching just one exponential moment using $\varphi(\alpha) \approx R^{\dlangle \tau \drangle}_{0,2}(\alpha)$.  The fit obtained with the MFPT plus two exponential moments ($m=0,\,n=3$) is practically indistinguishable from the exact $\varphi(\tau)$.  The next higher order approximations have imaginary poles. This provides an indication of overfitting and (correctly) suggests that the approximation should not be extended further.

While this procedure has worked on a few tested cases, as with all parameterized modelling approaches, success depends on a reasonable match between the form of the parameterized approximation and the true distribution. This can not be guaranteed but is a reasonable assumption since most FPTDs are expected to have distributions qualitatively like that shown in Fig.\ 2.

\section{Summary}
Mesoscopic coarse graining and the CTRW equations can be used to compute the macroscopic FPTD, $\varphi(\tau)$, of a complex stochastic system from short-term, and hence affordable, microscopic numerical simulations of its dynamics.  In many cases interest will focus on the MFPT and possibly a few additional low-order FPTD moments.  Instead of integrating the CTRW equations over time to compute $\varphi(\tau)$, a procedure that requires the full functional form of $K(\tau)$ to be estimated, and then integrating again to compute the moments, we have shown that the FPTD moments can be computed simply and directly from the moments of $\bar{K}$. This method is simpler and eliminates the quantization error inherent in the numerical solution of the CTRW equations in the time-domain. It can physically be viewed as an adaptation and extension of the steady-state flux-over-population method of computing transition rates, so we call it steady-state relaxation.

The steady-state expressions for the FPTD moments are simple enough for straightforward statistical error analysis, which permits the accuracy of the computed moments for a given amount of simulation data to be estimated.  This analysis can also be used to optimize the allocation of computational effort over the different mesoscopic states and to thereby reduce the total cost of the numerical simulations required for fixed accuracy. This is important since computability of the FPTD in large problems will often be limited by this cost. Such optimization improved efficiency over two-fold in a test problem with five mesoscopic states, and greater improvements are possible in problems with more states.  This improvement can be extended to the FPTD itself by modelling it using either a Gamma distribution or a rational-function approximation to its Laplace transform.

\section*{ACKNOWLEDGMENTS}
We are indebted to Ron Elber for may helpful discussions. T.F. was supported by NIH grant GM059796.

\appendix
\section{Bayes-Laplace Estimator}

In some cases (e.g., when numerical simulations are particularly costly), the goal may be just to estimate $\dlangle \tau \drangle$ to rough accuracy (e.g, 25--50\%) using the smallest possible number of simulations.  In such cases the $n_s$ may be small and there may be large fractional errors in the $\langle \bar{K} \rangle_{s's}$ whose effects are amplified by the matrix inversion.  Because the inversion is nonlinear, the maximum likelihood estimator used in \eqn{naive} may not be optimal and it is worth considering other possibilities.
One alternative is the Bayes-Laplace estimator \cite{Gelman:97}
\[
\langle \bar{K}\rangle_{s' s} \approx \frac{n_{s's}+1}{n_s+ \nu_s} \,,
\]
where $\nu_s$ is the number of states to which $s$ can make transitions. This is the mean Bayesian estimate of $\langle \bar{K}\rangle_{s' s}$ using a non-informative prior distribution (i.e., making the \emph{a priori} assumption that a system in state $s$ is equally likely to make a transition to any of the connected states $s'$). This estimator has a bias away from very small $\langle \bar{K}\rangle_{s's}$, suggesting that it may reduce the error of the inverted matrix. This surmise was empirically found to be true in the example of Sec.\ \ref{Sec:example}, but the improvement was only noticeable when the error was $>25\%$ \cite{used_Bayes}.

\newpage
\begin{center}
FIGURES
\end{center}
\includegraphics[width=7in]{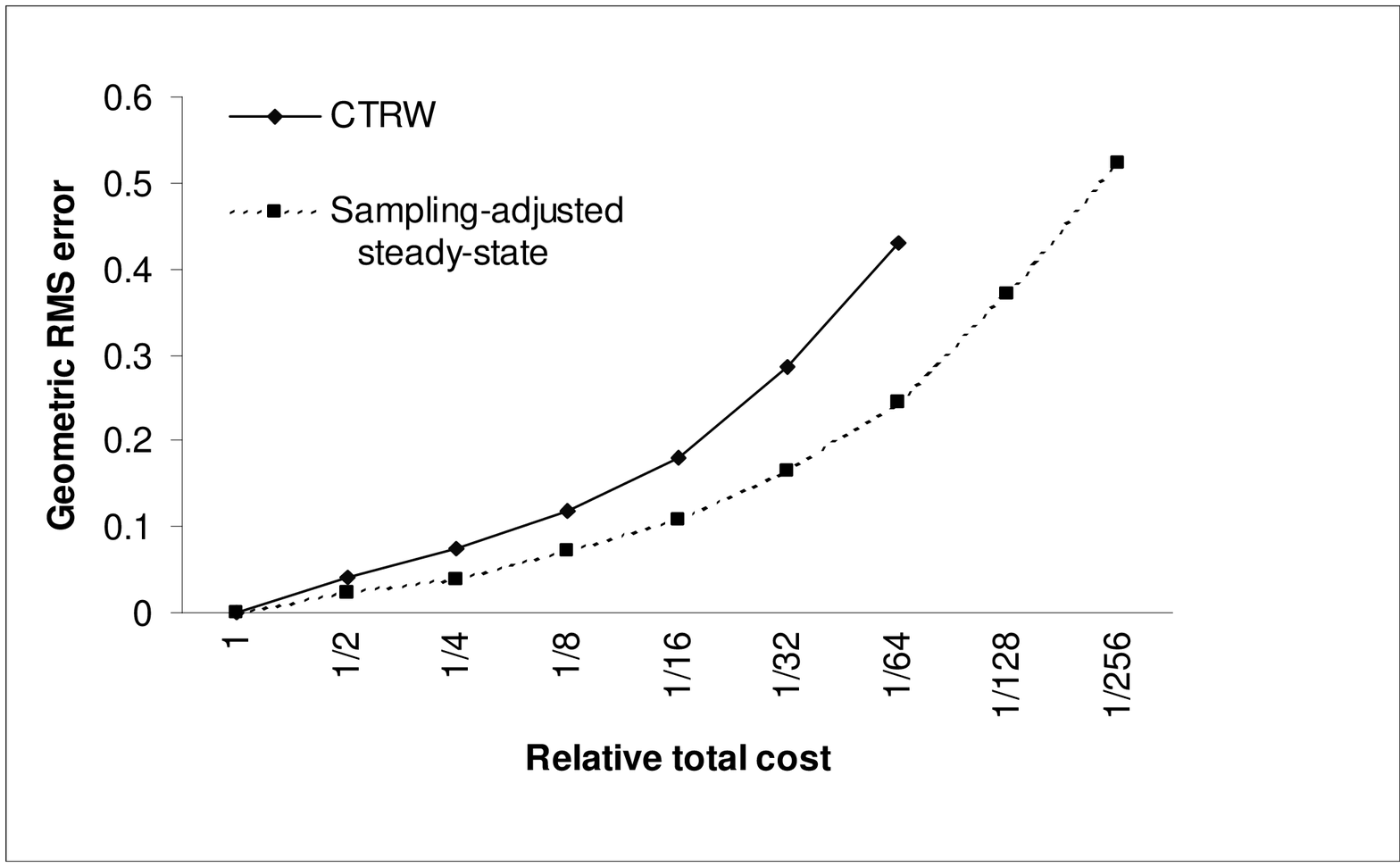}
\begin{figure}[ht!]
\caption{
Dependence of the geometric rms error of the calculated MFPT on the total cost of the simulation for the CTRW and sampling-adjusted steady-state methods applied to the two-dimensional entropic barrier model studied by Faradjian and Elber \cite{Faradjian:04}.  The error for the CTRW method can not be accurately estimated for $n_s=39$ and $n_s=19$ because too many trials have sparse patterns of transitions that fail to connect the initial and final states.}
\label{fig1}
\end{figure}

\begin{figure}[ht!]
\includegraphics[width=7in]{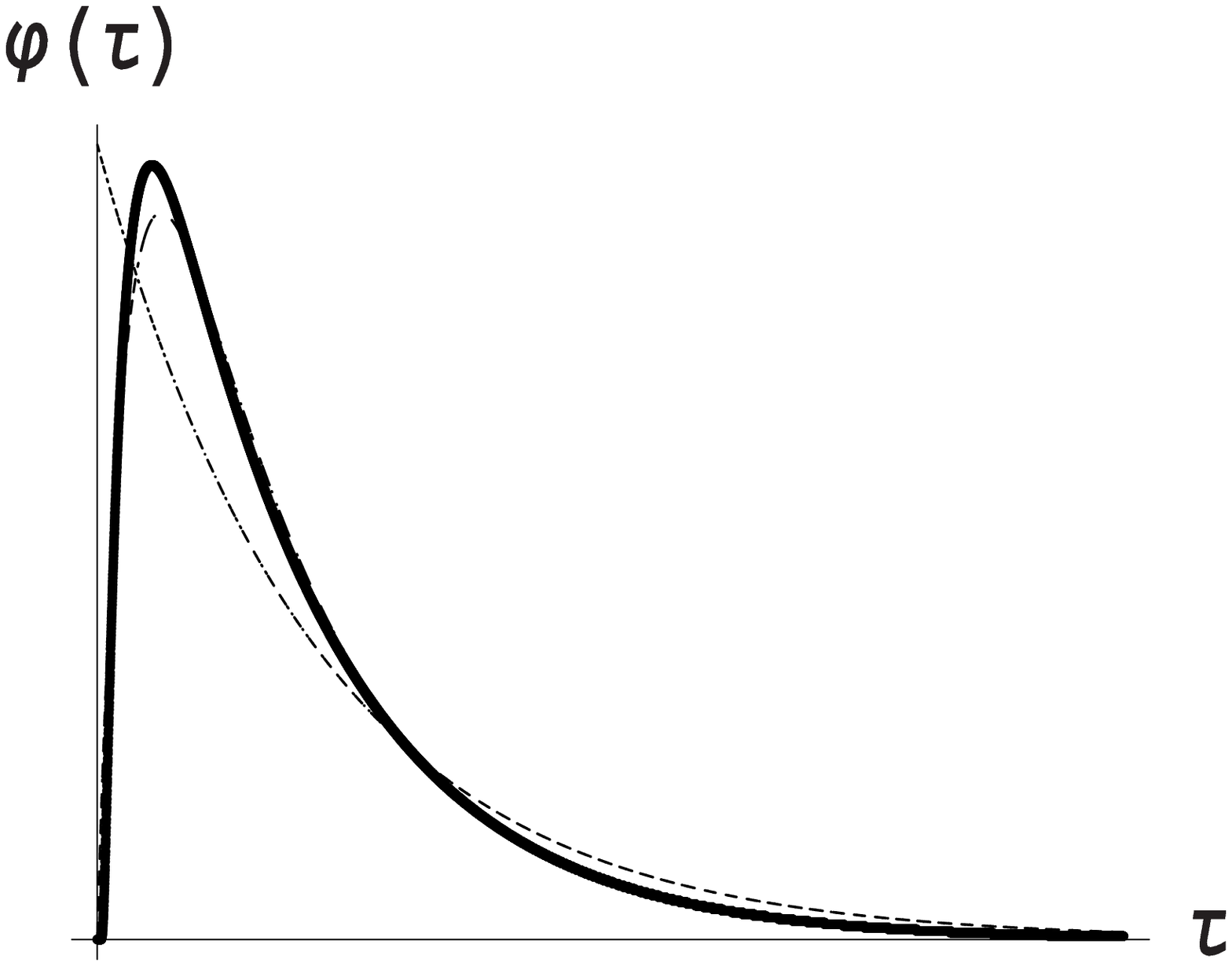}
\caption{
Comparison of the approximated and exact FPTDs for the two-dimensional entropic barrier model studied by Faradjian and Elber \cite{Faradjian:04}. The exact FPTD is displayed (solid line) along with the approximated FPTDs obtained by approximating the Laplace transform by a rational function matched to the MFPT alone (fine dotted line) or to the MFPT plus one (dashed line) or two (dotted line, indistinguishable from solid line) exponential moments.
}
\label{fig2}
\end{figure}


\begin{thebibliography}{99}
\bibitem{gme_note} The term ``generalized master equation'' has been used in multiple contexts. It is often used to describe the time-evolution of the quantum density matrix  \cite{VanKampen:92},  but here we are only interested in the classical form (i.e., corresponding to a diagonal density matrix). The form of the generalized master equation presented in Eqn.\ (8) of Ref.\ \onlinecite{Kenkre:74} only defines $P(t)$ for $t>0$. \eqn{generalized} here is a minor extension that defines $P(t)$ for all $t$ while being equivalent for $t>0$.
\bibitem{singular_note} Using \eqn{varphi_P} instead of \eqn{varphi_Pbar} would yield  $\tilde{\varphi}(-\alpha) = - \alpha \bfepsilon_f \cdot [\widetilde{\Gamma}(-\alpha) - \alpha I]^{-1} \cdot \bfepsilon_i$ instead of \eqn{Gamma_Laplace_relationship}. This expression has a removable singularity at $\alpha = 0$ that slightly complicates the analysis.
\bibitem{Elber_note} E.\ Vanden-Eijnden (unpublished result communicated by R.\ Elber) has noted that \eqn{K_Laplace_relationship} can also be derived by using the Laplace transforms of \eqns{QK} rather than the Laplace transform of \eqn{generalized}.
\bibitem{independent} $\phi_s$ is not fully independent from the  $p_{s'}^s$ because the mean waiting time for leaving $s$ may depend on the state to which a transition is made.  We can account for this by replacing $\phi_s$ in \eqn{variance} with $\sum_{s'} p_{s'}^s \phi_{s's}$, where $\phi_{s's} \equiv \langle \tau \bar{K} \rangle _{s's} \approx (n_{s's})^{-1} \sum_{i=1}^{n_{s's}} \tau^{s's}_i$ is the mean waiting time for the subset of first transitions out of $s$ that go to $s'$,  and replace \eqn{optimal} with
\begin{eqnarray*}
n_s & = & c \phi_s^{-1/2} \times \\
&& \sqrt{
\sum_{s' \ne s} \left(\pp{\dlangle \tau \drangle}{\phi_{s's}}\right)^2 \sigma^2_{\tau;s's} + \sum_{s'',s' \ne s }
\pp{\dlangle \tau \drangle}{p_{s''}^s} \sigma^2_{s;s'' s'}  \pp{\dlangle \tau \drangle}{p_{s'}^s} }\,.
\end{eqnarray*}
However, this requires estimation of the $\phi_{s's}$ and the variances of the $\{\tau^{s' s}_i\}$, $\sigma^2_{\tau;s's}$, which will require larger amounts of simulation data than estimation of $\phi_s$ and $\sigma^2_{\tau;s}$ alone.  We expect that this formula will be slightly better for optimizing the efficiency of high-accuracy results where enough simulations will be available to estimate the additional parameters, but it could be less effective than \eqn{optimal} for optimizing inexpensive low-accuracy results if not enough simulations were available to accurately estimate them. This is probably splitting hairs since we expect that, in most cases, overall error will be dominated by the error in estimating $\langle \bar{K} \rangle^{-1}$ which will not be affected by this change.
\bibitem{binomial_simplification} The sums over $s'$ and $s''$ need only include the states to which state $s$ makes transitions. In the special case where the states are connected in a linear order (i.e., $K$ is tri-diagonal), the expression can be simplified by replacing the multinomial parameters with a single binomial parameter for each $s \ne i,f$.
\bibitem{adaptive} In principle, a better procedure might be to adaptively adjust \eqns{optimal} as more samples are gathered. However, in the example of Sec.\ \ref{Sec:example} efficiency did not depend sensitively on the size of the pilot, so an adaptive procedure may not give much further improvement.
\bibitem{used_Bayes} When the maximum likelihood estimator of \eqn{naive} was used instead of the Bayes-Laplace estimator, the geometric rms error for $\texttt{cost}=\texttt{maxcost}/256$ increased from 52\% to 62\% and the error for $\texttt{cost}=\texttt{maxcost}/128$ increased from 37\% to 41\%.  The change in the other values was insignificant.
\bibitem{CTRW_computation} To reduce computational time and eliminate quantization errors, the CTRW values for infinitesimal quantization length were actually computed using the equivalent \eqns{K_first_moment} and \eqnref{estimating} with $n_s = n_{rm tot}/6$.
\bibitem{geometric_rms} $\dlangle \tau \drangle$ is a positive quantity and we expect errors in its calculation will be roughly lognormally distributed (this was empirically verified for the example), so  the rms geometric error is an appropriate error measure.  For exact value $\dlangle \tau \drangle$ and $N_{\rm tot}$ computed values $\dlangle \tau\drangle_i$ ($i = 1, \ldots, N_{\rm tot})$, the geometric rms error is $\exp[N_{\rm tot}^{-1}\sum_{i=1}^{N_{\rm tot}}\log^2(\dlangle \tau \drangle_i/\dlangle \tau \drangle)]-1$.
\bibitem{note2} $\langle \tau^k \varphi \rangle$ depends on $\langle \tau^k \bar{K}_D \rangle$ and $\langle \tau^{k'} \bar{K} \rangle$ ($0 \le k' < k$), so only one additional set of moments of $\bar{K}(\tau)$ needs to be computed for each additional FPT moment.
\bibitem{exponential_fit} We have found empirically that a slightly better fit for the same computational cost is obtained by determining $\beta$ and $\gamma$ by matching the MFPT and the second exponential moment $\langle \exp(- 2 \tau/\dlangle \tau \drangle) f \rangle = \dlangle \exp(- 2 \tau/\dlangle \tau \drangle) \drangle$, rather than using \eqns{Gamma_dist_fit}.  Moreover, this procedure is slightly simpler to use since it is not even necessary to expand \eqn{K_Laplace_relationship} to second order.
\bibitem{denser_interp} Since it is only the magnitude of $u_k$, not the number of interpolation points, that limits statistical accuracy, the interpolation points could be made denser with no significant increase in computational cost.  However, this is not expected to greatly improve accuracy.
\bibitem{constrained_rational} Matching the constrained rational function is numerically equivalent to matching an unconstrained rational function at the points $u = \{0, \delta, 2, 3, \ldots \}$, where $\delta$ is a small number.  Matching at $u=0$ ensures that $R_{m,n}^{\dlangle \tau \drangle}(0)= 1$ and matching at $u=\delta$ ensures \eqn{derivative}. This can be done using standard rational interpolation software or algorithms.
\end{thebibliography}
\end{document}